\documentclass{PoS}

\usepackage{epsfig}
\usepackage{amsmath}

\title{The proper motion and changing jet morphology of Cygnus X-3}

\ShortTitle{The proper motion of Cygnus X-3}

\author{\speaker{James Miller-Jones}\thanks{Jansky Fellow.}\\
        National Radio Astronomy Observatory, Charlottesville, USA\\
        E-mail: \email{jmiller@nrao.edu}}

\author{Charli Sakari\\
        Whitman College, USA\\
        E-mail: \email{sakaricm@whitman.edu}}

\author{Vivek Dhawan\\
        National Radio Astronomy Observatory, Socorro, USA\\
        E-mail: \email{vdhawan@nrao.edu}}

\author{Valeriu Tudose\\
        Netherlands Institute for Radio Astronomy, the Netherlands\\
        E-mail: \email{tudose@astron.nl}}

\author{Rob Fender\\
        University of Southampton, UK\\
        E-mail: \email{rpf@phys.soton.ac.uk}}

\author{Zsolt Paragi\\
        Joint Institute for VLBI in Europe, the Netherlands\\
        E-mail: \email{paragi@jive.nl}}

\author{Mike Garrett\\
        Netherlands Institute for Radio Astronomy, the Netherlands\\
        E-mail: \email{garrett@astron.nl}}

\abstract{We present analysis of 25 years' worth of archival VLA, VLBA and EVN
  observations of the X-ray binary Cygnus X-3.  From this, we deduce
  the source proper motion, allowing us to predict the location of the
  central binary system at any given time.  However, the line of sight
  is too scatter-broadened for us to measure a parallactic distance to the
  source.  The measured proper motion allows us to constrain the
  three-dimensional space velocity of the system, implying a minimum
  peculiar velocity of 9\,km\,s$^{-1}$.  Reinterpreting VLBI images
  from the literature using accurate core positions shows the jet
  orientation to vary with time, implying that the jets are oriented
  close to the line of sight and are likely to be precessing.}

\FullConference{Science and Technology of Long Baseline Real-Time Interferometry: \\
The 8th International e-VLBI Workshop, EXPReS09\\
		 June 22 - 26 2009\\
		 Madrid, Spain}

\begin{document}

\section{Introduction}

Cygnus X-3 is a high-mass X-ray binary system with a compact object
accreting matter from a Wolf-Rayet donor star.  The nature of the
compact object is still under debate.  The system is a persistent
radio source, with a typical flux density of 80--100\,mJy at GHz
frequencies.  Occasionally, it undergoes giant outbursts, in which the
radio flux density initially quenches to a level of a few mJy before
increasing by two to three orders of magnitude to peak at a level of
several Jy.  During such outbursts, VLBI observations have revealed
the presence of moving, relativistic jets (\cite{Mio01,Mil04,Tud07}),
predominantly aligned around a position angle $180^{\circ}$ E of N.
Hitherto however, analysis of the jet morphology has always been
hindered by uncertainty in identifying the location of the central
binary system; since VLBI observations were typically triggered on
detection of a radio flare, the jets were already resolved by the time
of the observations, often rendering an accurate identification of the
core position difficult or impossible from the images alone.

\section{Astrometric observations}

To resolve the uncertainty in identifying the core position, we
interrogated the NRAO archives for VLA and VLBA observations taken
during core-dominated epochs (i.e.\ outside major flaring episodes),
when the pointlike morphology of the source allowed us to accurately
identify the core position.  The VLA observations spanned the period
1983--2006, and we selected only data at frequencies of 8.4\,GHz or
higher, in the A-configuration of the array, to maximize the angular
resolution.  The VLBA observations were taken between 1997 and 2005.
We also considered EVN data taken by our group, using the e-VLBI
technique, between 2006 April and 2008 April, some of which has already
been published \cite{Tud07}.  Data were reduced according to standard
procedures in AIPS, with imaging and self-calibration being performed
using both AIPS and Difmap.

\begin{figure}
\centering
\epsfig{figure=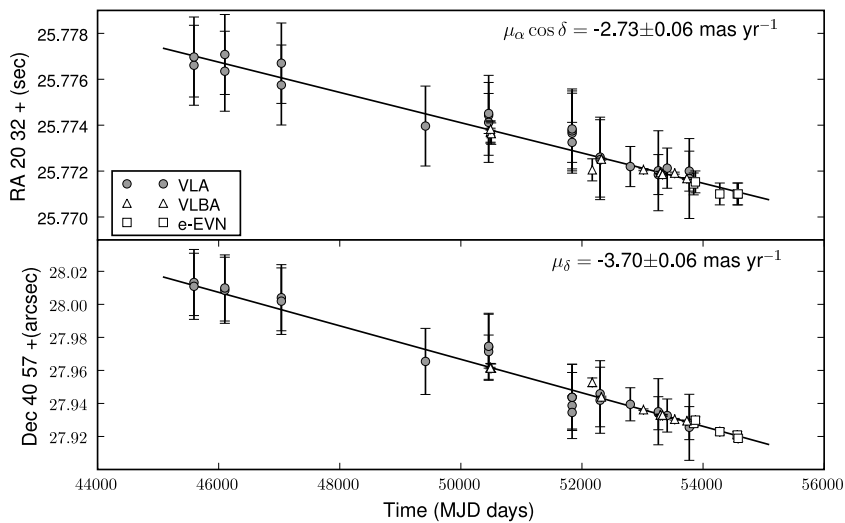,width=0.74\textwidth}
\caption{Variation in core position of Cygnus X-3 with time, in both
RA (top panel) and Dec (bottom panel).  The best-fitting proper
motions are shown by the solid lines.}
\label{fig:pms}
\end{figure}

Standard linear regression performed on the full dataset (see
Fig.~\ref{fig:pms}) then gives proper motions in Right Ascension and
Declination of
\begin{align}
\mu_{\alpha}\cos\delta &= -2.73 \pm 0.06 \textrm{ mas\,y}^{-1}\\
\mu_{\delta} &= -3.70 \pm 0.06 \textrm{ mas\,y}^{-1},
\end{align}
and a reference position on 2000 Jan 01 (MJD 51544.0) of 20$^{\rm
  h}$32$^{\rm m}$25$^{\rm s}$.7731(11), +40$^{\rm
  d}$57$^{\prime}$27$^{\prime\prime}$.951(12) (J2000).  Owing to the
  strong scattering along the line of sight to Cygnus X-3, it was not
  possible to determine the source positions sufficiently accurately
  to measure a parallactic distance to the source; the expected signal
  for a distance of 9--10\,kpc \cite{Dic83,Pre00} is of order
  0.1\,mas.

\subsection{Peculiar velocity}

Together with a line-of-sight systemic radial velocity and the
distance to the source, the proper motion can be used to derive the
full three-dimensional space velocity of the system \cite{Joh87}.
However, the radial velocity of Cygnus X-3 is very poorly constrained,
owing to the powerful stellar wind from the Wolf-Rayet donor star.
The high visual extinction along the line of sight forces us to rely
on infrared lines to measure the systemic velocity.  The P Cygni line
profiles from the outflowing wind, turbulence in the wind, and
uncertainty as to where in the system the observed infrared lines
arise, all hinder the determination of an accurate systemic radial
velocity.  \cite{Han00} constrain its magnitude to
$<200$\,km\,s$^{-1}$.  For a given three-dimensional space velocity,
we can derive the peculiar velocity of the source; the discrepancy
from what would be expected assuming the source participates in the
standard Galactic rotation.  Computing the peculiar velocity for the
full range of permitted systemic radial velocities (assuming a
Galactocentric distance of 8.4\,kpc and a Galactic rotational velocity
of 254\,km\,s$^{-1}$ \cite{Rei09}) gives a minimum peculiar velocity
of 9\,km\,s$^{-1}$ for a systemic radial velocity of
$-47$\,km\,s$^{-1}$.  For a 200\,km\,s$^{-1}$ systemic radial velocity
however, the peculiar velocity would be 250\,km\,s$^{-1}$.

The Wolf-Rayet nature of the donor star in Cygnus X-3 implies that the
system is young (typical Wolf-Rayet lifetimes are of order a few Myr).
It has therefore not had time to acquire a significant peculiar
velocity through assorted gravitational interactions during its
Galactocentric orbit, so the likelihood is therefore that any non-zero
peculiar velocity arises from a natal kick during the supernova in
which the compact object was formed.  However, with such an uncertain
systemic radial velocity, our limit on the peculiar velocity is not
terribly constraining.

For such a young system, there is a chance of tracing the
Galactocentric orbit of the system backwards in time to find the
birthplace of the binary.  However, owing to the large distance to the
source and its location behind one of the Galactic spiral arms, it was
not possible to identify a star cluster or supernova remnant from
which Cygnus X-3 might have originated.

\subsection{Implications for the jet morphology}

With the fitted proper motion of the system, we can identify the
position of the X-ray binary core at any given epoch.  This allows us
to more accurately interpret the observed jet morphology during
outbursts of the source.  We find that the jets were one-sided, and
flowing to the south during the 1997 February outburst, in agreement
with previous work \cite{Mio01}.  During the 2001 September outburst
\cite{Mil04}, they were two-sided and oriented north-south, but
brighter to the south, presumably owing to Doppler boosting of the
approaching jet.  However, during the 2006 May outburst, we find that
the jets were one-sided and flowing to the north.  In contrast with
the previous interpretation \cite{Tud07}, we are now able to identify
``knot C'' in their work as the true core.

The varying position angle of the jets supports the hypothesis that
the jets are precessing \cite{Mio01}, although the sampling is at
present too sparse to place useful constraints on the precession
period.  However, the dramatic position angle changes imply that the
jets are aligned very close to the line of sight, making Cygnus X-3
a good example of a Galactic microblazar.

\section{Conclusions}

We have measured the proper motion of Cygnus X-3 to be
$4.60\pm0.09$\,mas\,yr$^{-1}$.  From the proper motion and source
distance, we derive a minimum peculiar velocity of 9\,km\,s$^{-1}$, so
we cannot significantly constrain the size of any natal kick received
by the system when the compact object was formed.  The ability to
locate the central binary system at any given epoch allows us to
re-interpret archival VLBI images, showing that the position angle of
the jets can change by $180^{\circ}$ with time, supporting previous
evidence for jet precession and suggesting that the jets are oriented
very close to our line of sight.

\acknowledgments 

The VLA and VLBA are facilities of the National Radio Astronomy
Observatory, which is operated by Associated Universities, Inc., under
cooperative agreement with the National Science Foundation.  The
European VLBI Network is a joint facility of European, Chinese, South
African and other radio astronomy institutes funded by their national
research councils.

\end{document}